\documentclass[conference]{IEEEtran}
\usepackage{graphicx}
\usepackage{latexsym}
\usepackage{amsthm}
\usepackage{amsfonts}
\usepackage{subfigure}
\usepackage{dsfont}
\usepackage{epstopdf}
\usepackage{threeparttable}
\usepackage{multirow}
\usepackage{enumerate}
\usepackage{epsfig}
\usepackage{booktabs}
\usepackage{bm}
\usepackage{cite}
\usepackage{textcomp}
\usepackage{booktabs}
\usepackage{tabu}
\usepackage{color}
\usepackage{amsmath}
\usepackage{algorithm}
\usepackage{algpseudocode}
\newtheorem{remark}{Remark}
\usepackage{tabu}
\usepackage{booktabs}
\usepackage{diagbox}
\usepackage{bbding}
\newcommand{\tabincell}[2]{\begin{tabular}{@{}#1@{}}#2\end{tabular}}
\usepackage{makecell}
\usepackage{geometry}
\def\BibTeX{{\rm B\kern-.05em{\sc i\kern-.025em b}\kern-.08em
    T\kern-.1667em\lower.7ex\hbox{E}\kern-.125emX}}
\begin{document}

\title{Metric Learning-Based Timing Synchronization by Using Lightweight Neural Network\\
\thanks{This work is supported in part by the Sichuan Science and Technology Program (Grant No. 2023YFG0316, 2021JDRC0003), the Special Funds of Industry Development of Sichuan Province (Grant No. zyf-2018-056), and the Industry-University Research Innovation Fund of China University (Grant No. 2021ITA10016).}
}

\author{\IEEEauthorblockN{Chaojin~Qing$^\ast$, Na~Yang$^\ast$, Shuhai~Tang$^\ast$, Chuangui~Rao$^\ast$, Jiafan~Wang$^\ast$, and Hui Lin$^\dag$}
\IEEEauthorblockA{$^\ast$School of Electrical Engineering and Electronic Information,
Xihua University, Chengdu, 610039, China\\
%$^\dag$Synopsys Inc., 2025 NE Cornelius Pass Rd, Hillsboro, OR 97124, USA\\
$^\dag$ Hangtiankaite Electromechanical Technology Co., Ltd, Chengdu, 611730, China \\
Email: $^\ast$qingchj@mail.xhu.edu.cn, $^\dag$htktlh2016@163.com}}

\maketitle

\begin{abstract}
Timing synchronization (TS) is one of the key tasks in orthogonal frequency division multiplexing (OFDM) systems.
    However, multi-path uncertainty corrupts the TS correctness, making OFDM systems suffer from a severe inter-symbol-interference (ISI).
    To tackle this issue, we propose a timing-metric learning-based TS method assisted by a lightweight one-dimensional convolutional neural network (1-D CNN).
    Specifically, the receptive field of 1-D CNN is specifically designed to extract the metric features from the classic synchronizer.
     %of timing metric from the classic synchronizer.}
  %  the modal information of timing metric is first extracted from classic synchronizer.
%
%    a dedicated design of the receptive filed is proposed to easily learn the modal information of timing metric extracted by the algorithm of classic synchronizer.
    Then, to combat the multi-path uncertainty, we employ the varying delays and gains of multi-path (the characteristics of multi-path uncertainty) to design the timing-metric objective, and thus form the training labels. This is typically different from the existing timing-metric objectives with respect to the timing synchronization point. Our method substantively increases the completeness of training data against the multi-path uncertainty due to the complete preservation of metric information. By this mean, the TS correctness is improved against the multi-path uncertainty.
    %characteristics of CIR uncertainty (varying delays and gains of multi-path) to design the timing-metric objective, and thus form the training labels. With the
    %This forms a novel learning solution, which increases the completeness of training data against the CIR uncertainty. By this mean, the model's adaptability is enhanced against the CIR uncertainty, thus improving the TS correctness.
    Numerical results demonstrate the effectiveness and generalization of the proposed TS method against the multi-path uncertainty.
\end{abstract}


%\section{Introduction}
\section{Introduction}\label{I:I}
\IEEEPARstart{O}{rthogonal} frequency division multiplexing (OFDM) has been subject to extensive research efforts not only from the fifth generation (5G) systems but also from the Internet-of-Things (IoT) systems\cite{ref:IoT5GWiFi}.
In OFDM systems, a correct timing synchronization (TS) aims to find the starting of the receiver discrete Fourier transform (DFT) window within an inter-symbol-interference (ISI)-free region of an OFDM symbol\cite{ref:ISIf3}. Although synchronizing to this ISI-free region produces a phase rotation, this impairment can be easily countered by the channel equalization\cite{ref:ISI5}.
However, achieving this task is not easy due to the multi-path uncertainty.
The multi-path uncertainty is caused by the rich and diverse communication environments\cite{ref:CIRvar} and manifested in wireless channels with varying power delay profile (PDP).
%such as varying path gains and varying multi-path delays.
Because of the multi-path uncertainty, the timing metric is usually corrupted in non-light-of-sight (NLOS) scenarios. Consequently, the timing error, i.e., starting of receiver DFT window located outside the ISI-free region, is appeared, which will in turn affect the  subsequent signal processing.

To combat timing errors caused by the multi-path uncertainty, an alternative method for improving the TS correctness is to employ the joint mode, such as joint the TS and channel estimation, as done in\cite{ref:JSandCE}.
The method of joint TS and channel estimation\cite{ref:JSandCE} improves the TS correctness by partially counteracting the interferences of multi-path uncertainty. Nevertheless, this joint mode \cite{ref:JSandCE} results in a relatively high computational complexity.
Against the impairments caused by multi-path fading, noise, etc., an alternative method for improving the TS correctness is to deploy neural networks (NNs).
In this context, several machine learning-based studies have been conducted in finding high-performance TS methods for OFDM systems\cite{ref:CNNPD,ref:ELM-FTS,ref:ELM-labelTS}.
In \cite{ref:CNNPD}, a one-dimensional convolutional neural network (1-D CNN)-based TS method is investigated in OFDM systems, which improves the TS correctness relative to the conventional TS method.
Yet, this method ignores the impacts of multi-path uncertainty.
In \cite{ref:ELM-FTS}, the fine synchronization problem is investigated by assuming that the coarse TS and channel equalization have been achieved. Accordingly, \cite{ref:ELM-FTS} omits the consideration for multi-path interference, i.e., the multi-path uncertainty is neglected.
While the work in \cite{ref:ELM-labelTS} attempts to find ways to improve the TS correctness by designing training labels, the prerequisite of predicting the maximum multi-path delay limits its generalization performance.
To summarize, due to the lack of considering the high computational complexity in\cite{ref:JSandCE} and the multi-path uncertainty in\cite{ref:CNNPD,ref:ELM-FTS,ref:ELM-labelTS}, the machine learning-based TS for practical application is limited, inspiring us to investigate a lightweight machine learning-based TS method against the multi-path uncertainty.

In this paper, we propose a lightweight timing-metric learning-based TS method in OFDM systems.
To our best knowledge, against the multi-path uncertainty, the improvement of TS correctness by learning the timing metric has not been investigated.
The main contributions are listed as below.
\begin{itemize}
  %the receptive field of CNN layer is specially designed according to the finite lengths of channel impulse response (CIR) and cyclic prefix (CP). Meanwhile, the filter number are set based on the CP-based delay correlation. Also, an average pooling layer with patch equaling to the filter number is employed, which aims to reduce the data dimension sent for the fully connected layer. Furthermore, the hidden layer of fully connected layer is set to the length of the searching range of candidate timing offset.
  \item
  We propose the lightweight metric learning-based TS method.
  Different from \cite{ref:CNNPD}, the receptive field of 1-D CNN layer is specially designed and flexible according to the length of cyclic prefix (CP). Also, compared with \cite{ref:ELM-FTS,ref:ELM-labelTS}, the computational complexity of the designed neural network is significantly reduced.
  \item
  From the perspective of de-noising task, we specially design the timing metric to be learned. Specifically, the impact of uncertain multi-path delay on timing metric is considered to design the timing metric, and the impact of uncertain multi-path gain is also considered into the training stage. Thus, the adaptability of NN-based TS against multi-path uncertainty is improved.

\end{itemize}
\vspace{-2mm}
\begin{figure*}[t]
  \centering
  \includegraphics[width=0.85\textwidth]{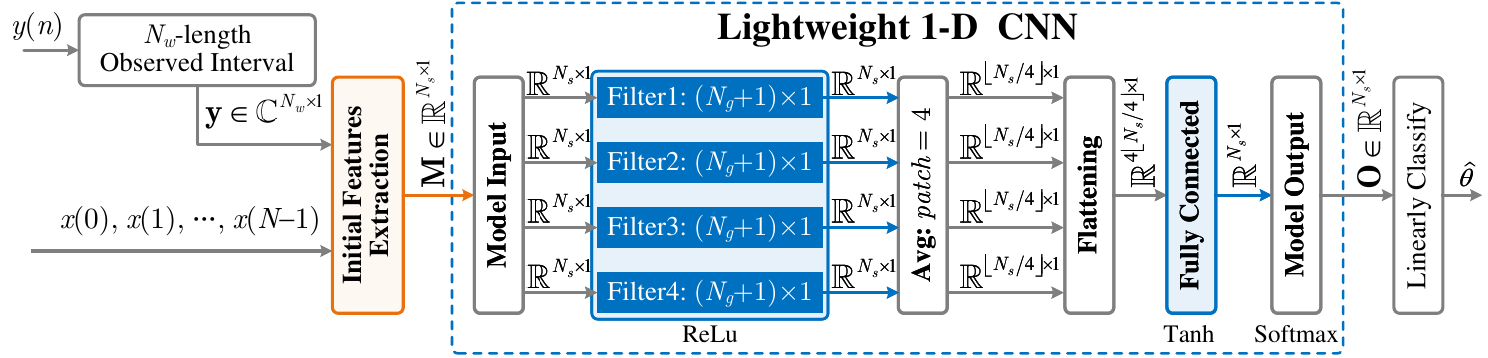}\\\vspace{-2mm}
  \caption{The proposed timing synchronizer.}\label{fig:sys}
\end{figure*}

\section{System Model and Problem Formulation}
\subsection{System Model}\label{S:II}
An OFDM system with $N$ sub-carriers is considered. At the transmitter, the time-domain OFDM symbol $\{s\left( n \right)\}^{N-1}_{n=0}$ is obtained by using the inverse DFT, i.e.,
\begin{equation}\label{EQ:Txdata}
s\left( n \right) = \sum\nolimits_{k = 0}^{N - 1} {S\left( k \right)e^ {j2\pi \frac{{kn}}{N}} },
\end{equation}
where $\{S(k)\}^{N-1}_{k=0}$ denotes the data/training symbol at the $k$th sub-carrier in the frequency domain.
$\mathbb{E}\{|s\left( n \right)|^2\}=P_t$ with $P_t$ being the transmitted power.

After appending the $N_g$-length cyclic prefix (CP), the transmitted signal consecutively passes through a multi-path fading channel.
With a $N_w$-length observed interval at the receiver, the received sample is expressed as
\begin{equation}\label{EQ:Rxdata}
y\left( n \right) = {{e^{\frac{{{j2\pi\varepsilon}n}}{N}}}\cdot\sum\nolimits_{l = 1}^L {h_l}{s\left( {n - {\tau _l}- \theta } \right)}}  + w\left( {n} \right),
\end{equation}
where $\varepsilon$ and $\theta$ respectively denote the normalized carrier frequency offset (CFO) and the unknown timing offset to be estimated. In \eqref{EQ:Rxdata}, $h_l$ and $\tau_l$ are the complex gain and normalized delay of the $l$th arriving path, respectively.
Meanwhile, $\tau_l=l-1$ and $0\le\tau_L<N_g$ are considered \cite{ref:Ch1}.
In \eqref{EQ:Rxdata}, $w(n)$ represents the complex additive white Gaussian noise with zero-mean and variance $\sigma_n^2$.

Then, the received $N_w$-samples $\{y\left( n \right)\}^{N_w-1}_{n=0}$ are buffered to form an observed vector $\mathbf{y}\in\mathbb{C}^{N_w\times1}$.
To observe at least one complete training sequence, $N_w\ge2N+N_g$ is required, and thus a discrete searching interval of unknown timing offset is employed, with its length being $N_s=N_w-N$.
In a classic synchronizer \cite{ref:CC}, the timing metric is utilized to estimate the unknown $\theta$. According to \eqref{EQ:Rxdata}, the timing metric, denoted as $\{M(d)\}^{N_s-1}_{d=0}$, is calculated as\cite{ref:CC}
\begin{equation}\label{EQ:TM}
M\left( d \right) = \frac{\left|\sum\nolimits_{k = 0}^{N - 1} {x^ *(k)}{y \left( {d + k} \right)}\right|^2}{\sum\nolimits_{k = 0}^{N - 1} \left|{y \left( {d + k} \right)}\right|^2},
\end{equation}
where $\{x(k)\}^{N-1}_{k=0}$ represents a local training sequence. Due to the impacts of multi-path fading, noise, etc., the metric in \eqref{EQ:TM} is easily impaired and then makes TS errors. Therefore, we develop a learning method to improve the TS correctness with lightweight network.
\vspace{-2mm}
\subsection{Problem Formulation}
Against the multi-path uncertainty, we focus on the improvements of the computational complexity and the adaptability of the deployed learning-based TS in OFDM systems.
Therein, $M(d)$ in \eqref{EQ:TM} can be regarded as the extracted initial feature, and then the impairments (e.g., noisy, multi-path interference) represented in $M(d)$ can be learned and remedied by NNs, as done in \cite{ref:ELM-labelTS}.
The de-noising problem can be mathematically formulated as
\begin{equation}\label{EQ:PBF}
\mathop {\min }\limits_{{\bf \Theta }} \left\| {{\mathbf{\Gamma}} - {G_{\bf \Theta} }\left( {\bf{M,{\Theta}}} \right)} \right\|_2^2,
\end{equation}
where $\bf\Theta$ is a set of network parameters to be optimized, and $G_{\bf\Theta}(\cdot)$ is a mapping function parameterized by $\bf\Theta$. In \eqref{EQ:PBF}, the vector forms that ${\mathbf{\Gamma}}=[\Gamma(0),\Gamma(1),\cdots,\Gamma(N_s-1)]^T$ and $\mathbf{M}=[M(0),M(1),\cdots,M(N_s-1)]^T$ denote the timing metric $\{\Gamma(d)\}^{N_s}_{d=1}$ to be learned and the initial feature $\{M(d)\}^{N_s}_{d=1}$ to be de-noised, respectively.
Nevertheless, the trained $G_{\bf\Theta}(\cdot)$ may suffer from a severe TS error due to the multi-path uncertainty. This is due to the fact that the multi-path interference represented in $\bf M$ is randomly unpredictable. Therefore, $\bf M$ are uncertain to be hardly recognized, degrading the correctness of learning-based TS in wireless propagation scenarios.
To handle this issue, the timing metric to be learned is specially designed to improve the TS correctness against multi-path uncertainty, which will be presented in \emph{Section III-B}.

\begin{table}
\vspace{-5mm}
\renewcommand{\arraystretch}{1.25}
\caption{Network Architecture}
\label{table_I}
\centering
\scriptsize
%\tiny
\setlength{\tabcolsep}{1.5mm}{
\begin{tabu}{c|c|c|c|c}
\tabucline[0.75pt]{-}
    Layer Name       & Output Size  &  Filter Size &Filter Number & Activation\\ \tabucline[0.75pt]{-}%\Xhline{0.8pt}
    Input Layer      & $(N_s,1,1)$  & -& - &-\\ \hline
     1-D CNN Layer    & $(N_s,1,4)$  & $N_g+1$ & 4 & ReLU \\ \hline
    Pooling \& Flattening  & $ (N_s,1)$  & -& -& -\\ \hline
    FC Layer  & $(N_s,1)$  & - & -& Sigmoid\\ \hline
    Output Layer & $(N_s,1)$  & -  & -&-\\
    \tabucline[0.75pt]{-}
\end{tabu}}
\vspace{-3mm}
\end{table}

\section{The Proposed Metric learning-based TS}
\subsection{Lightweight NN Architecture}
The proposed timing synchronizer is presented in Fig.~\ref{fig:sys} and summarized in \emph{TABLE I}, which consists of a classic correlator along with a NN process.
In the NN block, the single-layer 1-D CNN and single-layer fully connected NN are considered.
For 1-D CNN layer, the rectified linear unit (ReLU) is employed as the activation function.
As for fully connected layer, the tanh and softmax functions are employed in the hidden and output layers, respectively.

In the 1-D CNN block, the 1-D CNN deploys one convolution layer with 4 filters, and its the receptive field is selected as $(N_g+1)$.
Specifically, the receptive field of each filter layer is specially designed according to the finite lengths of channel impulse response (CIR) and CP.
This is due to the fact that the significant TS features are mainly appeared at arriving paths, and also the CIR length is less than the CP length. The TS feature extracted by using \eqref{EQ:TM} can be simplified as
\begin{equation}\label{EQ:TM2}
M\left( d \right) \approx \frac{{{{{P_t}} \mathord{\left/
 {\vphantom {{{P_t}} {\sigma _n^2}}} \right.
 \kern-\nulldelimiterspace} {\sigma _n^2}}}}{{1 + {{{P_t}} \mathord{\left/
 {\vphantom {{{P_t}} {\sigma _n^2}}} \right.
 \kern-\nulldelimiterspace} {\sigma _n^2}}}}\cdot\sum\nolimits_{l = 1}^L {{h_l}\delta \left( {d - {\tau _l} - \theta } \right)} .
\end{equation}
Since $N_g$ is usually less than one quarter of the symbol length $N$ (i.e., $N_g<0.25N$), the increase of computational complexity caused by a large receptive field can be alleviated.
Thus, $(N_g+1)$-size receptive field is suitable to capture the significant TS features extracted by the classic correlator.
Also, the number of filter  is set by considering that one complex multiplication (CM) equals 4 floating point operations (FLOPs), i.e., filter number is set as 4.
Thus, the total CMs of 1-D CNN processing approximately are equal to a CP-based correlation processing.

In the fully connected NN block, its hidden layer is selected according to the maximum searching length of candidate timing offset, i.e., $N_s$.
To further reduce the data dimension sent for the fully connected layer, an average pooling layer with patch equaling to the filter number is considered, i.e., 4-size patch.

In summary, the designed 1-D CNN and fully connected NN are constructed according to the parameters of $N_g$ and $N_s$, which are flexible in different scenarios. Meanwhile, by using CM as the evaluation of computational complexity, the computational complexity of the designed NN is $0.5N^2_s+N_sN_g$, while the correlation process in \eqref{EQ:TM} requires $N_sN$.
Since $N_s$ and $N_g$ are constrained by $N_s=N_w-N=N+N_g$ and $0<N_g<0.25N$, we will have $0.5N^2_s+N_sN_g-NN_s<0$.
Therefore, the designed NN is relatively lightweight compared with the classic correlator\cite{ref:CC}.
\vspace{-2mm}

\subsection{Timing Metric for TS Learning}
In the ISI-free region of per OFDM symbol, each sampling point can be regarded as the correct TS point\cite{ref:ISIf3}.
Consequently, the timing metric to be learned can be expressed as
\begin{equation}\label{EQ:Label1}
\Gamma \left( d \right) = \sum\nolimits_{\hat \theta  = {\theta+\hat\tau _L+1}}^{{\theta+N_g}} {\delta \left( {d - \hat \theta } \right)},
\end{equation}
where $ \hat \theta$ is the timing offset to be learned, and the $\hat \tau _L$ denotes the normalized maximum multi-path delay for offline training.
Usually, $\hat \tau_L$ is assumed to be fixed during the training stage.

However, due to the multi-path uncertainty, the real $\tau_L$ is unpredictable. For example, the root means square multi-path delay will change with time and propagation environments\cite{ref:WCPP}, making $\tau_L$ uncertain. Thus, it is highly possible that $\tau_L\neq\hat \tau_L$, resulting in an incorrect labeling.
%If the instances of incorrect labeling account for a certain number of the data set, the severe generalization error will appear, which significantly degrades the TS correctness.
When $\theta$ is fixed, the incorrect timing metric learned can be given by
\begin{equation}\label{EQ:ErrLabel1}
\begin{aligned}
\gamma \left( d \right) &= \sum\nolimits_{\hat \theta  = {\theta+{\hat \tau }_L+1}}^{{\theta+N_g}} {\delta \left( {d - \hat \theta } \right)} \oplus \sum\nolimits_{\hat \theta  = {\theta+\tau _L}+1}^{{\theta+N_g}} {\delta \left( {d - \hat \theta } \right)} \\
 &= \sum\nolimits_{\hat \theta  = {\theta+\tau _L+1}}^{{\theta+{\hat \tau }_L}} {\delta \left( {d - \hat \theta } \right)}.
\end{aligned}
\end{equation}
When $\hat{\tau}_L={\tau}_L$, the case of $\gamma \left( d \right)=0$ can be achieved, which means ideally labeling.
Due to the multi-path uncertainty, this case is hardly to be achieved.
Hence, we relax this demand by jointly considering these following motivations:
\begin{itemize}
  \item Although the cases of $\hat{\tau}_L\ge{\tau}_L$ make $\gamma \left( d \right)\neq0$, the set $\{\hat \theta\}^{\theta+\hat \tau_L}_{\theta+\tau_L+1}$ still belongs to the ISI-free region.
  \item Since ${\tau}_L$ is difficult to be predicted, ${\hat \tau}_L$ is no exception. Therefore, other priori information needs to be exploited for determining the value of ${\hat \tau}_L$.
  \item
      As NNs can compensate for deficiencies by learning from a certain number of data set, ${\hat \tau}_L$ for \eqref{EQ:Label1} can be expanded according to a set of random variables.
\end{itemize}
Given $N_t$-samples data set, we make $\{{\hat \tau}_{L,i}\}^{N_t}_{i=1}$ for \eqref{EQ:Label1} satisfy that ${\hat \tau}_{L,i}\mathop \sim \limits^{\textrm{i.i.d}}U[N_g/2, N_g-1]$.
By this mean, the timing metrics to be learned are expanded to increase the adaptability of NN against multi-path uncertainty.
\vspace{-1mm}
\begin{remark}
According to \eqref{EQ:ErrLabel1}, the main deficiency in \eqref{EQ:Label1} is caused by the dynamically changed ${\tau}_L$, resulting in error labeling and aggravating TS error.
Since learning models can compensate for deficiencies by learning from the prior inputs and objectives, the features to be learned can be expanded to increase the adaptability of NN against multi-path uncertainty.
To this end, the priori ${\hat \tau}_{L}\mathop \sim \limits^{\textrm{i.i.d}}U[N_g/2, N_g-1]$ is derived to expand the features of timing metrics.
Thus, the adaptability of trained model is enhanced against the uncertain ${\tau}_L$.
\end{remark}

In \emph{Section III-C}, the offline training and online deployment are described.
\vspace{-4mm}
\subsection{Offline Training and Online Deployment}
\subsubsection{Offline Training}
In this phase, $N_t=50,000$ is considered, which is split to the validation set and training set by 0.25.
The data set is denoted as $\{\mathbf{M}_i,{\bm\Gamma}_i\}^{N_t}_{i=1}$, in which $\mathbf{M}_i$ is obtained via \eqref{EQ:Txdata}--\eqref{EQ:TM} and ${\bm\Gamma}_i$ is obtained by using \eqref{EQ:Label1}.
Therein, ${\hat \tau}_{L,i}\mathop \sim \limits^{\textrm{i.i.d}} {U}[\lfloor N_g/2\rfloor, N_g-1]$ is utilized to alleviate the effect of uncertain multi-path delay. An exponentially decayed channel model \cite{ref:Ch1} with decayed exponent $\eta$ is considered. Meanwhile, $\eta_i\mathop \sim \limits^{\textrm{i.i.d}}U(0.01, 0.2)$ is employed to alleviate the effect of uncertain multi-path gains.
Besides, $\theta_i\mathop \sim \limits^{\textrm{i.i.d}}{U}\left[0,N-1\right]$.

For the designed NN in \emph{TABLE I}, optimizer employs the stochastic gradient descent (SGD) algorithm, and its initial learning rate is set as $\alpha=0.002$ \cite{ref:adam}.
By respectively denoting $B$ and $J$ as the batch size and the number of steps, the network optimization is defined as\cite{ref:adam}
\begin{equation}\label{EQ:SGD}
  {{{\bf \Theta} _{q + 1}} \leftarrow {{\bf \Theta}_q} - \alpha \nabla \frac{1}{B}\sum\nolimits_{i = (q-1)B+1}^{qB} {\left\| {{G_{{\bf \Theta}_q} }\left( {{{\bf{M }}_i}},{{\bf \Theta}_i} \right) - {{\bf{\Gamma}}_i}} \right\|_2^2}} ,
\end{equation}
where the subscript $q$ denotes the $q$th iterative step for optimizing, and $1\le q\le J$.
\subsubsection{Online Deployment}
By using \eqref{EQ:Txdata}--\eqref{EQ:TM}, ${M}(d)$ is obtained, forming $\mathbf{M}=[M(0),M(1),\cdots,M(N_s-1)]^T$.
Then, with the optimized $G_{\bf \Theta}(\cdot)$, the model output, denoted as ${\bf O}\in \mathbb{R}^{N_s\times1}$, is given by ${\bf O} = G_{\bf\Theta}\left({\bf M}\right)$. Finally, by expressing $\bf O$ as $[O(0),O(1),\cdots,O(N_s-1)]^T$, the estimated timing offset is
\begin{equation}\label{EQ:EstTO}
  \widehat{\theta}  = \mathop {\arg \max }\limits_{0 \le d \le {N_s} - 1} \left\{ {{O(d)}} \right\}.
\end{equation}

In \emph{Section IV-B} and \emph{Section IV-C}, the effectiveness and generalization of the proposed TS method against the multi-path uncertainty are presented.
\begin{table}
\vspace{-4mm}
\renewcommand{\arraystretch}{1.25}
\caption{Abbreviations of Different TS Methods}
\label{table_III}
\centering
\scriptsize
%\tiny
\setlength{\tabcolsep}{0.5mm}{
\begin{tabu}{c|l}
\tabucline[0.75 pt]{-}
 Abbreviation      & Computational Complexity (CM)   Example\\
 \tabucline[0.75 pt]{-}%\Xhline{0.8pt}
    ``Prop''               & \ \ The proposed method             \\ \hline
   ``Prop without $\Gamma$'' &  \tabincell{l}{\ \  The proposed method directly learns the received signal\vspace{-0.5mm}\\without the initial feature extraction} \\ \hline
   ``Prop with fixed $\hat \tau_L$''      & \tabincell{l}{\ \ The proposed method does not specially design the timing\vspace{-0.5mm}\\metric for model training, i.e., fix $\tau_L=22$ for training}        \\ \hline
   ``Ref\cite{ref:JSandCE}''      & \ \ The joint TS and channel estimation method in \cite{ref:JSandCE}\\ \hline
    ``Ref\cite{ref:ELM-labelTS}''& \ \ The label designed-based ELM method in \cite{ref:ELM-labelTS} \\ \hline
    ``Ref\cite{ref:CTS}''      & \ \ The classic TS method proposed in \cite{ref:CTS}\\ \hline
    ``DNN'' & \tabincell{l}{\ \ A conventional back-propagation NN which owns two hidden\vspace{-0.5mm}\\dense layers, with both neuron nodes being $N_s$}\\
    \tabucline[0.75 pt]{-}
\end{tabu}}
\end{table}
\section{Simulation Results}\label{S:IV}
In the simulations, we consider basic parameters as that $N=128$, $N_g=\lfloor N/4\rfloor=32$\cite{ref:ELM-FTS}, $N_w=2N+N_g=288$, and $N_s=N_w-N=160$.
%Besides, the probability of missing the TS point of ISI-free region is hereinafter named as the probability of TS error, which is utilized to evaluate the performance of TS correctness.
For simulated channel models, they have not been utilized for offline training.
For the sake of clarity, we list the abbreviations of different timing synchronization methods in \emph{TABLE II}.
\begin{table}
\vspace{-4mm}
\renewcommand{\arraystretch}{1.5}
\caption{Computational Complexity among Different TS Methods}
\label{table_II}
\centering
\scriptsize
%\tiny
\setlength{\tabcolsep}{3.5mm}{
\begin{tabu}{c|c|c}
\tabucline[0.75 pt]{-}
 Method      & Computational Complexity (CM)  & Example\\
 \tabucline[0.75 pt]{-}%\Xhline{0.8pt}
    Ref\cite{ref:JSandCE}      & $LN{N_s} + \sum\nolimits_{l = 1}^L {( {3l{N_s} + {l^3} + {l^2}{N_s}} )}$              &  $1371536$             \\ \hline
    Ref\cite{ref:ELM-labelTS}   & $16N_s^2+ 4{N_s}+1.5N -4$              &  $410428$     \\ \hline
    {DNN}      & $0.75N_s^2+ N{N_s}+2N_s+N-2$              &  ${40126}$ \\ \hline
    {Proposed}      & $1.5N_s^2+ 3{N_s} +N-2$              &  ${39006}$            \\
    \tabucline[0.75 pt]{-}
\end{tabu}}
\vspace{-3mm}
\end{table}
\vspace{-4mm}
\subsection{Computational Complexity}\label{VB}
The comparison of computational complexity among different TS method is illustrated in \emph{TABLE~\ref{table_II}}. Therein, the total channel paths are selected as 23, i.e., $L=23$, and other parameters are adopted from \emph{Section IV-A}.
According to \emph{TABLE~\ref{table_II}}, ``Prop'' reaches the smallest CM  among the given TS methods.
Therefore, ``Prop'' has the superiority in realizing lightweight network.\vspace{-2mm}
\subsection{Effectiveness Analysis}
\begin{figure}[t]
\vspace{-3mm}
  \centering
  \includegraphics[width=0.5\textwidth]{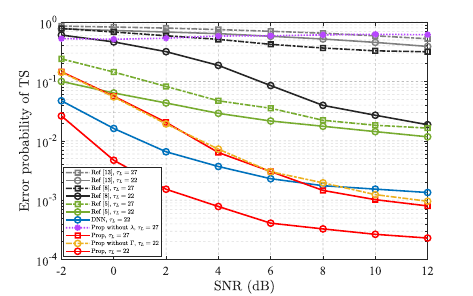}\\\vspace{-3mm}
  \caption{Effectiveness in NLOS scenarios.}\label{fig:eff}\vspace{-4mm}
\end{figure}
To analyze the effectiveness, Fig.~\ref{fig:eff} depicts the error probability of TS. Wherein, an un-trained exponential decayed factor that $\eta=-\ln(10^{-\frac{10}{10}})/(L-1)$\cite{ref:Ch1} is utilized and the maximum multi-path delay changes from $22T$ to $27T$, which are utilized to simulate the multi-path uncertainty.
In Fig.~\ref{fig:eff}, for each given value of $\tau_L$, the probability of TS error for ``Prop'' is smaller than those of ``Ref\cite{ref:CTS}'', ``Ref\cite{ref:JSandCE}'', and ``Ref\cite{ref:ELM-labelTS}''. Meanwhile, for all given SNRs, ``Prop'' achieves a lower probability of TS error than ``DNN''. This is because CNN is easier to capture data features compared with DNN methods. It is noteworthy that, although $\tau_L$ increases from $22$ to $27$ caused by the multi-path uncertainty, ``Prop'' exhibits slight generalization error than ``Prop with fixed $\hat\tau_L$'', due to the use of the designed timing-metric objective.
Besides, ``Prop'' reaches a smaller TS error than ``Prop without $\Gamma$'', which demonstrates the benefits of learning timing-metric.
In summary, the performance improvements of ``Prop'' is effective against the multi-path uncertainty.\vspace{-2mm}
\subsection{Generalization Analysis}
\begin{figure}[t]
\vspace{-1mm}
  \centering
  \includegraphics[width=0.5\textwidth]{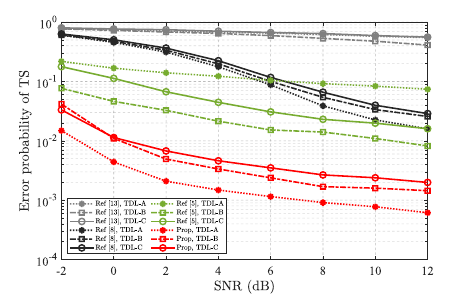}\\\vspace{-3mm}
  \caption{Generalization in NLOS scenarios.}\label{fig:gen}\vspace{-3mm}
\end{figure}
Fig.~\ref{fig:gen} plots the comparison of the error probability of TS to analyze the generalization performance of ``Prop'' against different 5G tapped-delay-line (TDL) channel models\cite{ref:3GPP5G}.
Notably, these channel models have not been used for offline training.
For each given channel model, ``Prop'' achieves smaller probability of TS error among the given TS methods in the whole SNR region.
Besides, for ``Prop'', the fluctuation in the probability of TS error caused by different un-trained channel models are not obvious.
Therefore, the proposed TS method (i.e., ``Prop'') has a good generalization capability against different 5G TDL channel models.
\vspace{-2mm}
\section{Conclusion}\label{S:V}
In this paper, we investigate a lightweight timing-metric learning-based TS in OFDM systems, which alleviates the multi-path uncertainty by utilizing the designed timing-metric objective.
%Specifically, the developed synchronizer learn the timing metrics, and this improves the TS correctness from the classic synchronizer.
%Especially, by designing the timing-metric objective, the generalization performance of the proposed network is enhanced against the CIR uncertainty.
%Meanwhile, the proposed method realizes lightweight network, compared with the jointly processing method and ELM-based methods.
Different from \cite{ref:JSandCE,ref:CNNPD,ref:ELM-FTS,ref:ELM-labelTS}, against the multi-path uncertainty, we utilize the proposed lightweight network along with the designed learning solution to learn the timing metric, which improves the TS correctness and generalization performance with less computational complexity.
By simulations, numerical results exhibit the superiority of the proposed method in reducing the error probability of TS against multi-path uncertainty, whilst revealing its good generalization performance against different un-trained 5G TDL channel models.
\section{Acknowledgment}
This work is supported in part by the Sichuan Science and Technology Program (Grant No. 2023YFG0316, 2021JDRC0003), the Special Funds of Industry Development of Sichuan Province (Grant No. zyf-2018-056), and the Industry-University Research Innovation Fund of China University (Grant No. 2021ITA10016).

\ifCLASSOPTIONcaptionsoff
\newpage
\fi
\vspace{-4mm}
\bibliographystyle{ieeetran}
\bibliography{ref}
%\bibitem{b1} G. Eason, B. Noble, and I. N. Sneddon, ``On certain integrals of Lipschitz-Hankel type involving products of Bessel functions,'' Phil. Trans. Roy. Soc. London, vol. A247, pp. 529--551, April 1955.
%\bibitem{b2} J. Clerk Maxwell, A Treatise on Electricity and Magnetism, 3rd ed., vol. 2. Oxford: Clarendon, 1892, pp.68--73.
%\bibitem{b3} I. S. Jacobs and C. P. Bean, ``Fine particles, thin films and exchange anisotropy,'' in Magnetism, vol. III, G. T. Rado and H. Suhl, Eds. New York: Academic, 1963, pp. 271--350.
%\bibitem{b4} K. Elissa, ``Title of paper if known,'' unpublished.
%\bibitem{b5} R. Nicole, ``Title of paper with only first word capitalized,'' J. Name Stand. Abbrev., in press.
%\bibitem{b6} Y. Yorozu, M. Hirano, K. Oka, and Y. Tagawa, ``Electron spectroscopy studies on magneto-optical media and plastic substrate interface,'' IEEE Transl. J. Magn. Japan, vol. 2, pp. 740--741, August 1987 [Digests 9th Annual Conf. Magnetics Japan, p. 301, 1982].
%\bibitem{b7} M. Young, The Technical Writer's Handbook. Mill Valley, CA: University Science, 1989.
%\end{thebibliography}
%\vspace{12pt}
%\color{red}
%IEEE conference templates contain guidance text for composing and formatting conference papers. Please ensure that all template text is removed from your conference paper prior to submission to the conference. Failure to remove the template text from your paper may result in your paper not being published.

\end{document}